\documentclass[preprint2]{aastex}
\usepackage{epsfig}
\usepackage{amsmath}

\shorttitle{Direct Distances}
\shortauthors{Homan and Wardle}

\begin{document}

\title{Direct Distance Measurements to Superluminal Radio Sources}

\author{D. C. Homan\altaffilmark{1} and J. F. C. Wardle\altaffilmark{2}}
\affil{Physics Department, Brandeis University}
\affil{Waltham, MA 02454} 
\affil{ }
\affil{Accepted for publication in The Astrophysical Journal}
\altaffiltext{1}{dch@quasar.astro.brandeis.edu}
\altaffiltext{2}{jfcw@quasar.astro.brandeis.edu}

\begin{abstract}
We present a new technique for directly measuring the distances to 
superluminal radio sources.  By comparing the observed proper motions
of components in a parsec scale radio jet to their measured Doppler factors, we
can deduce the distance to the radio source independent of the standard rungs
in the cosmological distance ladder.  This technique requires
that the jet angle to the line of sight and the ratio of pattern to flow velocities
are sufficiently constrained.  We evaluate a number of possibilities for
constraining these parameters and demonstrate the technique on a well defined
component in the parsec scale jet of the quasar 3C\,279 (z = 0.536).  We find 
an angular size distance to 3C\,279 of greater than $1.8^{+0.5}_{-0.3}\eta^{1/8}$ 
Gpc, where $\eta$ is the ratio of the energy density in the magnetic field to the 
energy density in the radiating particles in that jet component.  For an 
Einstein-de Sitter Universe, this measurement would constrain the Hubble constant 
to be $H_0 \lesssim 65\eta^{-1/8}$ km/s/Mpc at the two sigma level.  Similar 
measurements on higher redshift sources may help discriminate between cosmological
models.
\end{abstract}


\section{Introduction}

To answer fundamental questions about the geometry of the Universe and 
the distribution of matter on the largest scales we must be able to measure
the distance to objects at large redshift.  Type Ia supernovae have begun
to fill this role and recent results 
suggest that the overall expansion of the Universe may be accelerating 
\citep{R98,P99}.  
The great brightness of Type Ia supernovae, which can out-shine
their parent galaxies, allow them to be observed up to redshifts of $z=1$.  
It is  important to have cosmological probes at even higher redshifts where
we are less sensitive to local perturbations \citep{T99} and the 
predictions of different cosmological models become more distinct \citep{CPT92}. 

Quasars are regularly observed at very large redshifts and provide an
alternative to gravitational lenses for high redshift cosmology studies.
Optically, it is difficult to define a standard candle for quasars because 
of the wide range of intrinsic luminosities between objects and the high degree of 
variability of individual objects.  Similar difficulties exist with using the 
radio luminosity for the $\sim$ 10\% of objects that are radio loud.  
Global Very Long Baseline Interferometry (VLBI) arrays offer an alternative to
defining a standard candle.  The ability to resolve compact structures and observe 
proper motions in the radio jets of many sources suggests the possibility 
of defining a standard rod from which angular size distances may be derived.

\citet{K93} defined a standard rod for compact radio sources as
the distance between the core and the most distant jet component whose
peak brightness exceeds 2\% of the core brightness.  He used data from
82 compact sources (out to $z \approx 3$) and found an observed relation
between angular size and redshift which is consistent with an 
Einstein-de Sitter Universe (deceleration parameter, $q_0 = 0.5$).

\citet{LB77} and \citet{LBL78} combined
early proper motion measurements with a light-echo model, 
e.g. \citep{C39}, in the first attempts to use proper motions observed in
radio jets as cosmological distance indicators.  
Several authors have examined the statistics of proper motions 
of patterns in radio jets as a cosmological probe.  
\citet{Y79} proposed the idea of a
proper motion-redshift diagram to measure Hubble's constant
and the deceleration parameter.  
\citet{CBPZ88} used measured proper motion 
data and a simple beaming model to show that the upper envelope of
the proper motion data decreased with redshift in a manner consistent
with a Friedmann cosmology and inconsistent with several alternatives.
More recently, \citet{V94} have refined these ideas
and used a much larger sample to explore the possibility of
measuring cosmological parameters with proper motion data.  They found 
they could usefully constrain cosmological parameters and simultaneously
learn about the distribution of jet parameters. 

Our approach differs from these techniques in that we use VLBI observations to
directly measure the distance to {\em individual} superluminal radio
sources.  We compare the proper motions of individual components in a parsec-scale
radio jet with measurements of their Doppler factors.  In addition, we must
constrain the angle the component motion makes with the line
of sight and separate the pattern speed (observed in proper
motion measurements) from the flow speed (observed in Doppler
factor measurements).  In \S{2} we present a number of possibilities for
constraining these parameters, and in \S{3} we evaluate the technique with an
example, using a well defined component in the VLBI jet 
of the quasar 3C\,279.  Section \ref{s:sys-err} 
discusses the measurement errors and systematic 
uncertainties associated with this technique. Section \ref{s:cosmo} explores 
the application of such measurements to cosmological questions.  
Our conclusions are presented in \S{\ref{s:conclude}}.

\section{Theory}

The observed proper motion, $\mu$, of a pattern (component) in a parsec-scale  
radio jet depends on the intrinsic pattern speed, $\beta_p$, 
the angle between the jet axis and the line to the observer, $\theta$, and the
angular size distance to the radio source, $D_A$.
\begin{equation}
\label{e:mu}
\mu = \frac{c}{D_A(1+z)}\frac{\beta_p\sin\theta}{1-\beta_p\cos\theta} 
    = \frac{c}{D_A(1+z)}\beta_a
\end{equation}
\noindent where $\beta_a$ is the apparent transverse velocity of the 
component in units of the speed of light.

We define a Doppler factor for the pattern, $\delta_p$, by
\begin{equation}
\label{e:d}
\delta_p = \frac{\sqrt{1-\beta_p^2}}{1-\beta_p\cos\theta}.
\end{equation}

The aberrated angle, $\theta'$, is the angle between the jet axis and the line
to the observer in the frame that moves along the jet with the pattern speed. 
It is given by
\begin{equation}
\label{e:cos-theta'}
\cos\theta' = \frac{\cos\theta-\beta_p}{1 - \beta_p\cos\theta} 
            = \cos\theta - \beta_a\sin\theta
\end{equation}
\noindent and
\begin{equation}
\sin\theta' = \delta_p\sin\theta.
\end{equation}

\begin{figure}
\epsfig{file=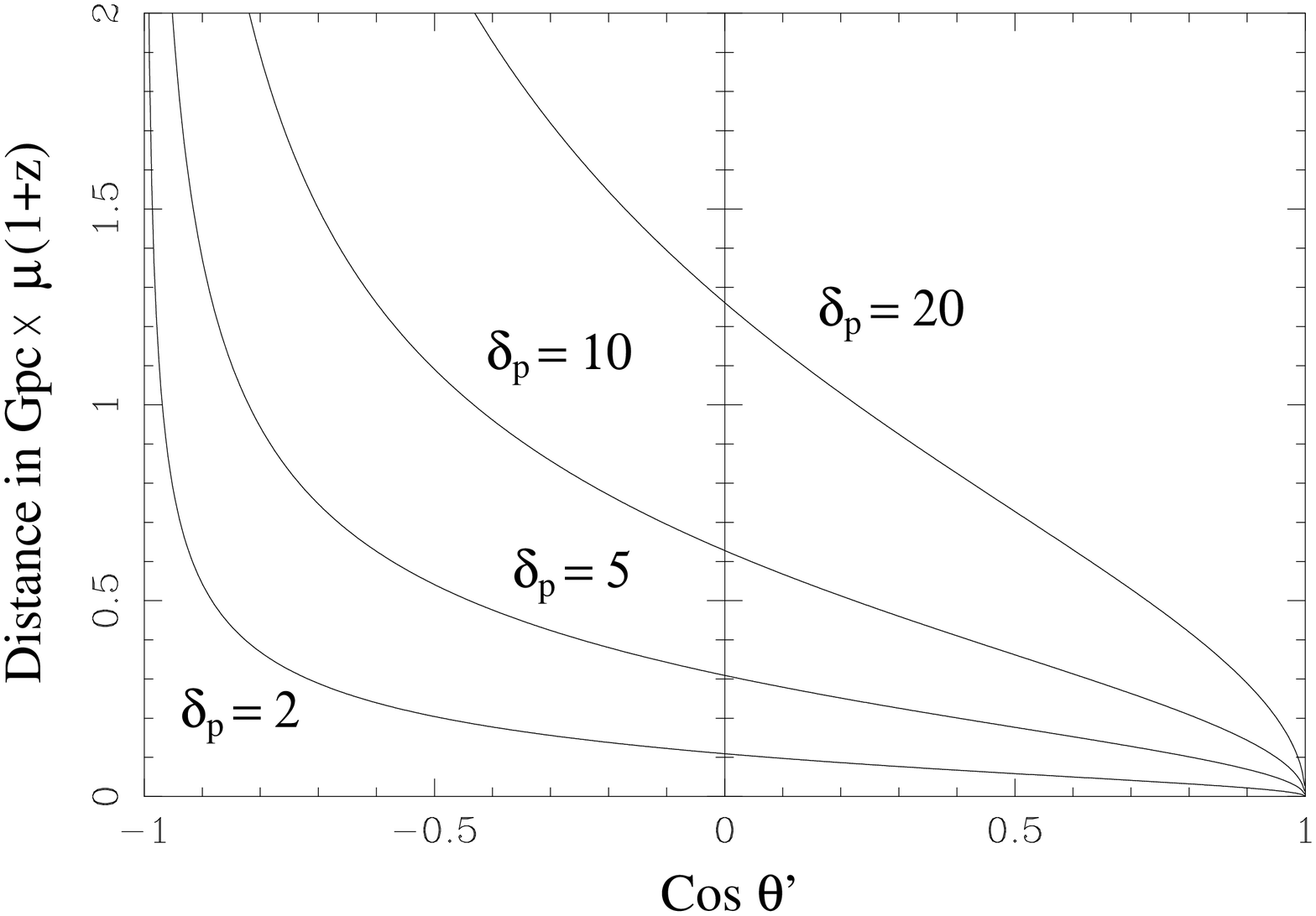,width=4in}
\caption{\label{f:dist-theta}
A plot of angular size distance in Gpc$\times\mu(1+z)$ versus $\cos\theta'$ 
where $\theta'$ is the angle of observation in the pattern frame and $\mu$ is the
observed proper motion in mas/yr. The very
large distances near $\cos\theta' = -1$ are a curious feature of this plot.
Patterns with a negative $\cos\theta'$ are viewed from behind and travel 
at a larger angle than the critical angle, $\beta = \cos\theta_c$.  A source 
which has a high Doppler factor, $\delta_p$, and a negative $\cos\theta'$ 
must have a very small critical angle, $\theta_c$.  With a very small critical
angle, such a source has a very large $\beta$ giving it a large apparent
speed, $\beta_a \propto D_A\times\mu$.} 
\end{figure}

We can use these relations to derive an 
expression for the angular size distance: 
\begin{equation}
\label{e:dist-theta}
D_A = \frac{c}{\mu(1+z)}\left(
\frac{\sqrt{\delta_p^2+\cos^2\theta'-1}-\delta_p\cos\theta'}{\sqrt{1-\cos^2\theta'}}
\right).
\end{equation}
\noindent Equation \ref{e:dist-theta} is plotted for various values of $\delta_p$ in 
figure \ref{f:dist-theta}.  In this equation $\mu$ is in natural units of rad/sec; however,
in figure \ref{f:dist-theta} $\mu$ is in units of mas/year. 

\placefigure{f:dist-theta}
 
The proper motion $\mu$ is a directly observed quantity. The Doppler factor of 
the pattern, $\delta_p$, is indirectly observed through the use of Synchrotron 
Self-Compton (SSC) or equipartition arguments (see \S{\ref{s:dops}} and the appendix) 
which measure a product of Doppler factors:
\begin{equation}
\delta_{SSC} = \delta_p(\delta'_f)^\frac{2\alpha+3}{2\alpha+4}
\end{equation}
\noindent and
\begin{equation}
\label{e:delta_f'}
\delta_{eq} = \delta_p(\delta'_f)^\frac{2\alpha+5}{2\alpha+6}
\end{equation}
\noindent where $\alpha$ is the spectral index ($S \propto \nu^{-\alpha}$) for optically
thin synchrotron radiation and $\delta'_f$ is the Doppler factor of
the fluid frame relative to the pattern frame:
\begin{equation}
\delta'_f = \frac{\sqrt{1-(\beta'_f)^2}}{1-\beta'_f\cos\theta'}
\end{equation}
\noindent where $\theta'$ is the angle of observation in  
the pattern frame and $\beta'_f$ is the speed of the fluid
in the pattern frame.

For comparison to proper motion measurements, we are interested 
in $\delta_p$.  To determine $\delta_p$ from equipartition or SSC
arguments we must measure or usefully constrain $\delta'_f$ which
typically requires measuring the pattern versus flow speed;  however,
there is a useful constraint
we can place on $\delta'_f$ if we know $\theta'$.  For a given $\theta'$, 
the maximum in $\delta'_f$ is when $\beta'_f = \cos\theta'$, so
\begin{equation}
\label{e:limit-df}
\delta'_f \leq 1/\sqrt{1-(\cos\theta')^2}.
\end{equation}

\subsection{Additional Constraints}
\label{s:constrain}

In the sections that follow, we explore a number of constraints that
allow us to turn measurements of proper motions and Doppler factors into direct
distance measurements.  Some of these techniques allow measurement 
or constraint of $\delta'_f$ which is important for accurate determination
of $\delta_p$ from Doppler factor measurements.

\subsubsection{Bent Jets}

For a given pattern speed, $\beta_p$, the observed proper motion is maximized when 
$\cos\theta = \beta_p$.  At this critical angle, $\cos\theta' = 0$ and
equation \ref{e:dist-theta} reduces to:
\begin{equation}
D_A = \frac{c\sqrt{\delta^2_p-1}}{\mu(1+z)}
\end{equation}

Sources inside the critical angle will give an upper limit on $D_A$, and 
sources outside the critical angle will give a lower limit.  However, 
these limits will only be useful for sources near the critical angle, and 
for these sources, we need an additional constraint to determine
their orientation relative to the critical angle.  

Jets which bend on VLBI scales give a unique opportunity for observing
a source at or near its critical angle.  As a component on a curved trajectory
passes through the critical angle, a number of observable
effects occur: the proper motion of the component maximizes, the orthogonal
component of the magnetic field (projected in the plane of the sky) maximizes
creating a maximum or minimum in the observed linear polarization,
and thin features, such as shocks, minimize in observed aspect ratio.  

We note that if the flow and pattern have different speeds 
they will also have different critical angles.  The maximization of the
proper motion occurs at the pattern's critical angle.  Any 
maximum or minimum in the observed linear polarization depends on the
flow's critical angle.  Observing both critical angles provides a method
of resolving the difference between the flow and pattern speeds.  If only
the critical angle of the pattern is observed, a useful constraint is
that the Doppler factor for the flow as observed from the pattern 
frame, $\delta'_f$, has a maximum value of 1 (see equation \ref{e:limit-df}).  

\subsubsection{Aspect Ratio}

Several authors have made use of a sharp (narrow) feature in a radio jet to 
measure or constrain the jet angle to the line of sight \citep{ES83,BOH83,BOC89,UW92}.
A sharp feature, assumed to be oriented perpendicular to the jet direction, is a sign that
the pattern is moving at close to its critical angle.
The shape an observer sees for a component moving in a radio jet is governed
by the aberration between the pattern frame and observer frame. 

\begin{figure}
\epsfig{file=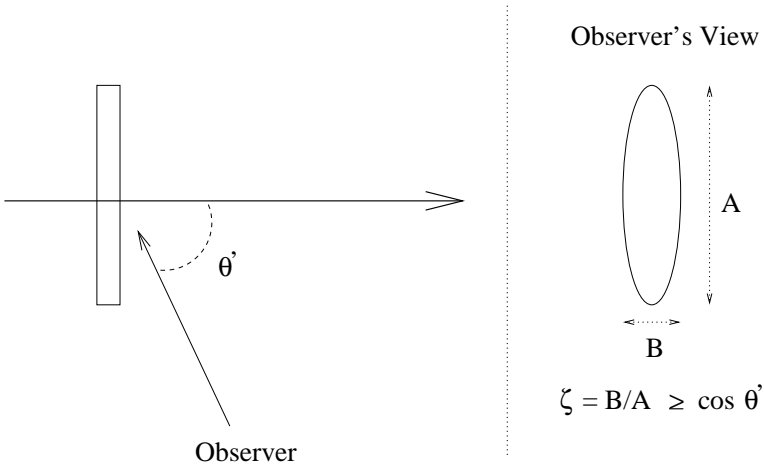, width = 4in}
\caption{\label{f:aspect}
An observer sees the profile of an axially symmetric
pattern from the aberrated angle, $\theta'$.  The measured aspect ratio, $\zeta$,
is always greater than or equal to $\cos\theta'$.}
\end{figure}

\placefigure{f:aspect}

The observed aspect ratio of a component is the ratio of its extent along
the jet to its extent transverse to the jet , $\zeta = size_\|/size_\bot$ (see
figure \ref{f:aspect}). The ratio $\zeta$ 
constrains $\cos\theta'$ if the pattern is assumed to be axially symmetric and 
oriented perpendicular to the direction of motion.  Under these circumstances, 
$\zeta \geq |\cos\theta'|$ and we obtain potentially useful limits on 
the angular size distance (from equation \ref{e:dist-theta}, corresponding to
positive and negative signs for $\cos\theta'$):
\begin{equation}
\label{e:lower}
D_A \geq \frac{c}{\mu(1+z)}\left(
\frac{\sqrt{\delta_p^2+\zeta^2-1}-\delta_p\zeta}{\sqrt{1-\zeta^2}}
\right)
\end{equation}
\noindent and
\begin{equation}
\label{e:upper}
D_A \leq \frac{c}{\mu(1+z)}\left(
\frac{\sqrt{\delta_p^2+\zeta^2-1}+\delta_p\zeta}{\sqrt{1-\zeta^2}}
\right)
\end{equation}

Because the SSC technique produces only a lower limit on the Doppler
factor \citep{M87}, only the lower limit on $D_A$ will be applicable
when we determine $\delta_p$ using that technique.   The equipartition 
assumption will also produce a lower limit on the Doppler factor if we only
have an upper limit on the frequency of the self-absorption turnover in 
the synchrotron spectrum \citep{R94}. 

It is important to note that the relations developed here assume that 
the pattern is perpendicular to the jet direction and not oblique.
Obliqueness in the plane of the sky is observable from the orientation
of the feature and perhaps its linear polarization.  Assuming it is not
very large, this kind of obliqueness can be corrected for;
however, if a component can be oblique in the plane of 
the sky it may also be oblique in the plane of observation.  Obliqueness
in the plane of observation is indistinguishable from effects of aberration
for determining the observed component dimensions and may cause the 
relation $\zeta \geq |\cos\theta'|$ to be violated.  Uncertainty in 
the degree of obliqueness of a given component is equivalent to an added
uncertainty in the measurement of $\zeta$ \citep{BOC89,UW92}.

\subsubsection{Linear Polarization}

The linear polarization of a pattern in a radio jet provides a measure
of the magnetic field order.  For a tangled magnetic field which has 
been compressed (due perhaps to a propagating shock), the degree of
linear polarization observed depends both upon the degree of 
compression and the viewing angle \citep{L80,HAA85}.
The highest degrees of parallel linear polarization (for a given compression)
will be observed when the flow is moving at or near its critical 
angle.  The degree of compression can be related to the speeds
of the flow (upstream and downstream) relative to the propagating
shock, e.g. \citep{CW88,HAA89}.

\citet{W94} work out a complete model for deducing 
the jet angle to the line of sight and pattern versus flow speeds
from detailed VLBI polarization data.  They consider the general
case of a compression in a jet with a tangled field plus a component
of ordered field along the axis of compression.  By measuring the
degree of linear polarization and total intensity in both the shocked 
and un-shocked regions in the jet of 3C\,345 (z = 0.595), they were 
able to constrain the flow speed relative to the shock and 
the inclination of the jet to the line of sight. 

To illustrate the use of the constraints available from linear
polarization observations, we will start with the results of \citet{W94}
for jet component C3 of 3C\,345. The reader is
referred to that paper for details.  For their 1984.2 epoch, 
they find nominal values of $\theta = 2.5^\circ/D_A$, $\beta_d = 0.5$, and 
$\beta_a = 11.1D_A ~~(\mu=0.44$ mas/year).
The flow speed of the shocked fluid towards the core in the frame of 
the shock-front is $\beta_d$.  In the notation of this paper, 
$\beta_f' = -\beta_d$.  $D_A$ is measured in Gpc for the values given 
above.  

We use equations \ref{e:cos-theta'} and \ref{e:delta_f'} to calculate
$\cos\theta' = 0.5$ and $\delta_f' = 0.7$ which are essentially independent 
of distance.  If we had observations of the synchrotron self-absorption 
turnover for component C3 in 3C\,345 at this epoch, we could measure
its total Doppler factor using equipartition or SSC arguments.  We could
then use the measurement of $\delta_f'$ to determine the Doppler factor 
of the pattern, $\delta_p$.  With $\delta_p$, $\mu$, and $\cos\theta'$ 
determined, equation \ref{e:dist-theta} would allow calculation of the distance 
to 3C\,345.  

\subsubsection{Jet/Counter-Jet Ratio}

Another potentially useful constraint is the observed jet/counter-jet brightness
ratio, $R$, e.g. \citep{UW92}.
\begin{equation}
\label{e:jetratio}
R = \left(\frac{1+\beta\cos\theta}{1-\beta\cos\theta}\right)^{n+\alpha}
    = (\beta_a^2 + \delta^2)^{n+\alpha}
\end{equation}
\noindent where $n=3$ for discrete components and $n=2$ for continuous jet
emission.  The reduction of the equation to include $\beta_a$ is only valid
if the pattern and flow speeds are the same.  If they are the same or if we 
know the relationship between them, measurement of $R$ allows us to directly
compare the Doppler factor to observed proper motion and deduce the distance
to the source.  In the event that the pattern and flow speeds are the same, 
the angular size distance is given by
\begin{equation}
D_A = \frac{c\sqrt{R^{\frac{1}{n+\alpha}} - \delta^2_p}}{\mu(1+z)}
\end{equation}

An attractive feature of this approach is that the final answer
depends weakly on the measurement of $R$.  For highly beamed sources, however,
$R$ is huge and even high quality VLBI measurements cannot usefully
constrain it.  For less beamed sources, measuring or constraining $R$
is more promising.  A major drawback to using $R$ to connect 
$\mu$ and $\delta_p$ is that $R$ is a global property of the jets
rather than of an individual component.  For $R$ to be useful, the flow speed and
angle to the line of sight need to be the same for the jet and counter-jet
and constant over the region for which $R$ is measured.  In addition, there cannot
be significant excess absorption of emission from the counter-jet, c.f. \citep{K98}.  

\subsection{Doppler Factors}
\label{s:dops}
The synchrotron spectral turnover provides
a kind of natural (but broad) spectral line for homogeneous synchrotron sources.
By carefully measuring the spectrum of a component and its angular size we
can use limits on its SSC x-ray flux, e.g. \citep{M87} or
an assumption of equipartition between the field and particle energies \citep{R94}
to determine the Doppler factor.  In the appendix these formulae are 
presented for arbitrary homogeneous geometry and for the specific case
of a spherical geometry.  

We have chosen to use a spherical component geometry for the calculations presented 
in this paper. Without detailed knowledge of the true geometry a 
spherical geometry is well-suited for calculation because it computes 
the angular area presented to the observer reasonably, 
gives a sensible line of sight through the component, and provides 
naturally for a range of optical depths across the component.  
\citet{M87} suggests using $\theta_d \simeq 1.8\sqrt{\theta_{G_a}\theta_{G_b}}$ 
to convert Gaussian FWHM dimensions (measured in model-fitting) to spherical diameters, 
and we adopt his approximation. (See appendix \ref{s:model-geom} for discussion
of the effect of assumed model geometry on measured Doppler factor.)

\section{Example: 3C\,279}

As an example of these ideas, we use a well defined component in the
milli-arcsecond jet of the well known blazar 3C\,279 at z = 0.536.  We 
observed 3C\,279 with the Very Long Baseline Array\footnote{The VLBA is part of
the National Radio Astronomy Observatory, which is a facility of the National Science
Foundation operated under cooperative agreement by Associated Universities, Inc.}
(VLBA) for six epochs at 15 and 22 GHz during 1996 and at four frequencies 
(5.0, 8.4, 15.4, and 22.2 GHz) during December of 1997.  These observations
were all calibrated using standard techniques, e.g. \citep{C93,RWB94}, using the 
National Radio Astronomy Observatory's Astronomical Image Processing
System (AIPS).  Model-fitting was performed in the (u,v)-plane 
with the Caltech VLBI program, DIFMAP.

\subsection{Proper Motion}

\begin{figure}
\epsfig{file=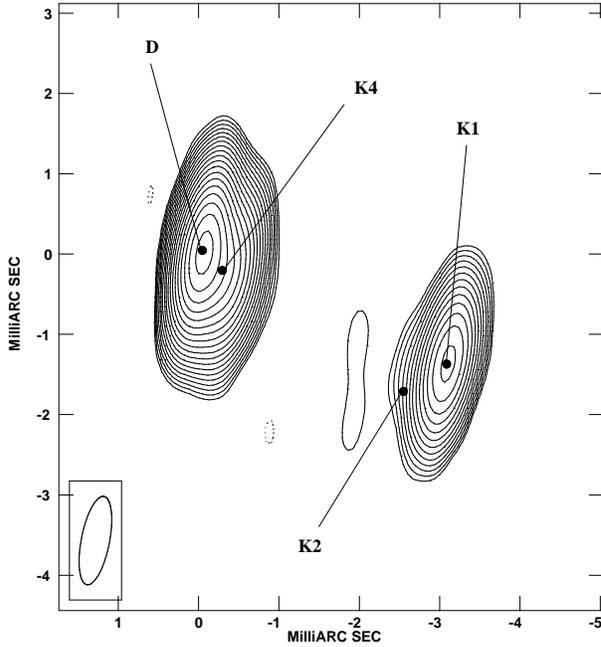, width=4in}
\caption{\label{f:3C279K}
Naturally weighted image of the jet of 3C\,279 at 22 GHz.  Epoch 1997.94.
The locations of the components K1, K2, K4, and D are marked on the image. 
Component data are summarized in table \ref{t:3c279t}.
}
\end{figure}

\begin{figure}
\epsfig{file=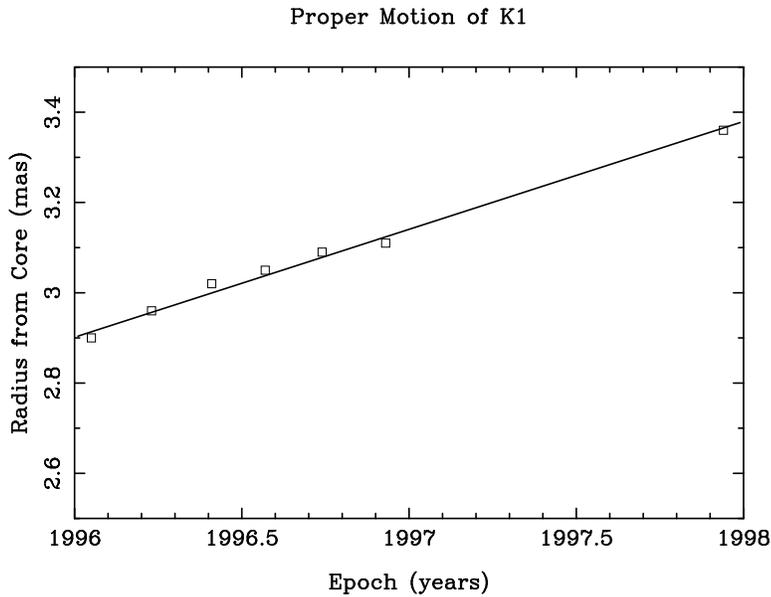,angle = -90,width=4in}
\caption{\label{f:3C279-mu}
Proper motion of component K1.  Component positions are taken from the 15 GHz
model-fits.  With the data weighted equally, the derived proper motion is
$\mu = 0.24 \pm 0.01$ mas/year.  Error bars on the positions are not plotted
because we do not have a good a-priori method for estimating them.  From the
small deviation of the data from the fit, we can say that the errors in the radial
position of K1 are $\sim 0.02$ which is about $1/20$ th of the uniformly 
weighted beam width along that direction.  
}
\end{figure}

Figure \ref{f:3C279K} shows the structure of the inner jet of 3C\,279 at
22 GHz.  Table \ref{t:3c279t} gives detailed component
data from model-fitting the inner jet of 3C\,279 for the 1997.94 epoch.
K1 is a well defined, strong component which has persisted for
years, e.g. \citep{U98}.  The component is located approximately
3 milli-arcseconds from the core at a position angle of $-115^\circ$.  Over
the course of our observations we observe this component to move radially
from the core with a proper motion of $\mu = 0.24 \pm 0.01$ mas/year (see
Figure \ref{f:3C279-mu}). It maintains a structural position angle of 
$-114^\circ \pm 1^\circ$ over the course of our observations. 

\placefigure{f:3C279K}

\placetable{t:3c279t}

\placefigure{f:3C279-mu}

\subsection{Synchrotron Self-Absorption\\ Turnover}

\begin{figure}
\epsfig{file=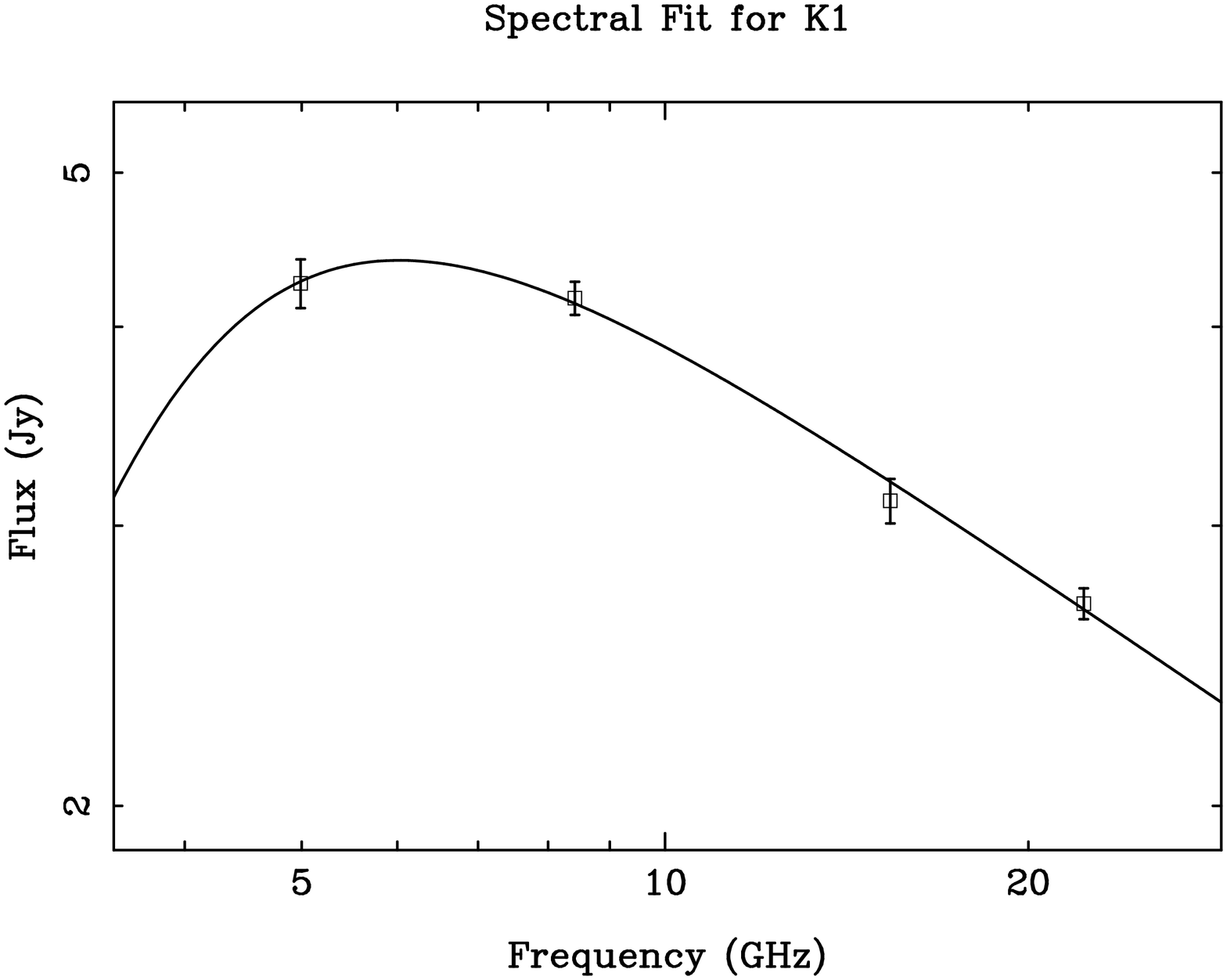,angle = -90,width=4in}
\caption{\label{f:spec.k1}
Synchrotron self-absorption spectrum of component K1. $\nu_{peak} = 6.02$ GHz $(+0.33,
-0.49)$, 
$S_{peak} = 4.40$ Jy $(+0.19,-0.07)$, and $\alpha = 0.52 \pm 0.05$.   
}
\end{figure}

We fit the total intensity of K1 at all four frequencies 
in 1997.94 and have measured its spectral turnover.
Figure \ref{f:spec.k1} displays the fit of a synchrotron self-absorption
spectrum to the data assuming a slab geometry.
(Fitting the spectral shape of a homogeneous sphere gives a nearly identical
result but a slightly smaller error range on the parameters.)  The spectral 
turnover is located at $\nu_{peak} = 6.02$ GHz $(+0.33, -0.49)$ with a 
flux, $S_{peak} = 4.40$ Jy $(+0.19,-0.07)$, and
a spectral index, $\alpha = 0.52 \pm 0.05$.  The errors in the fit are
approximately 1 $\sigma$ errors found by a Monte Carlo simulation.  The simulation
created and fit 1000 fictional data sets using the measured data and assuming the
measurements are Gaussian distributed with 1 $\sigma$ deviations given by
the measured error bars.  The error bars on the fluxes were estimated by
varying parameters in the model-fits using Jim Lovell's {\em Difwrap} program,
an interactive shell for the Caltech VLBI program, DIFMAP.  A number of 
factors were used to gauge the size of the error bars including shape
of the chi-square minimum, noise on the residual map, and direct
comparison of model and data in the (u,v)-plane.  We have no direct way
of knowing if the errors estimated for the fluxes are genuinely 
1 $\sigma$ errors; however, the spectral fit has 1 degree of freedom (4 data points
and 3 parameters) so the $\chi^2$ of the spectral fit should be near unity if the
errors on the data are 1 $\sigma$.  The measured $\chi^2$ of the 
spectral fit is 1.0.  

\placefigure{f:spec.k1}

The chief uncertainty in the measured fluxes of K1 is due the presence
of its poorly defined ``tail'', fit as component K2.  While K1 is fit 
robustly by a sharp Gaussian component, K2 is more difficult to
fit.  This becomes more of a problem at the lower frequencies 
(especially 5 GHz) where K1 is not as well resolved.  

\begin{figure}
\epsfig{file=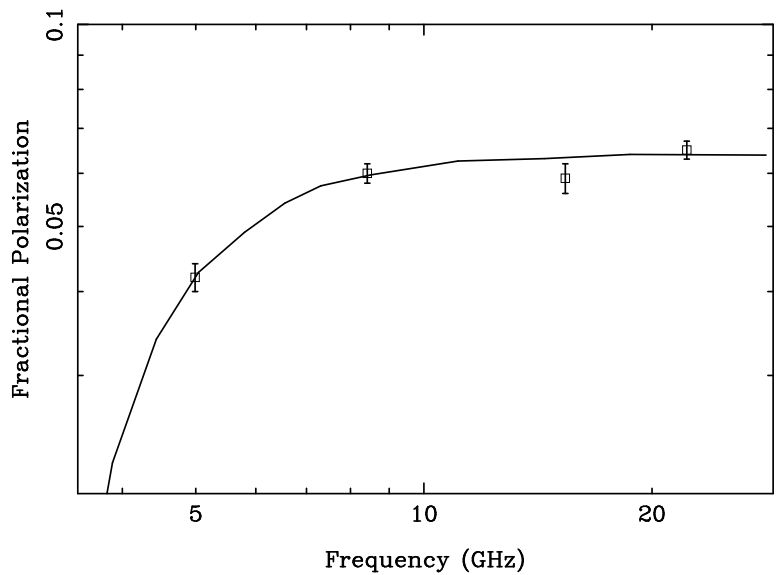,angle = -90,width=4in}
\caption{\label{f:k1.lp}
Fractional linear polarization of K1 plotted against frequency.  The
``theoretical curve'' is from a numerical simulation of a homogeneous
slab with a completely tangled magnetic field plus a small ordered 
component.  The size of the ordered component was scaled to
match the observed fractional polarization at 22 GHz.  The opacity
of the slab was fixed to match the fit to the total intensity spectrum.
}
\end{figure}

To check the spectral fit for K1, we examined the spectrum of observed
{\em fractional} linear polarization.  Figure \ref{f:k1.lp} displays
the observed fractional polarization plotted together with a theoretical curve
produced by numerical simulation.  The simulation is of a homogeneous slab
with the same total intensity spectrum as fit to K1.  The simulation
solved the full equations of polarized transfer, e.g. \citep{JO77}, for
a completely tangled magnetic field plus a small ordered component.  The
magnitude of the ordered component was scaled to give the 
observed fractional polarization at 22 GHz.  To simplify the simulation, 
no internal Faraday rotation was allowed. It is clear that the spectrum of the 
fractional polarization is completely consistent with total intensity 
spectrum fit to K1.  

\placefigure{f:k1.lp}

\subsection{Doppler Factor from Equipartition}
\label{s:calc-dopp}
We now use the measured angular size of K1 at 22 GHz and the fit to the
spectral turnover to deduce an equipartition Doppler factor (derived in
the appendix).  The measured FWHM angular size at 22 GHz is 
$\theta_{Ga}\times\theta_{Gb} = 0.46(\pm 0.02)\times0.20(\pm 0.01)$ mas$\times$mas. 
The error bars were estimated by varying parameters in the model-fit and 
by comparison to the measurements at 15 GHz and 8 GHz.  
We use the Doppler factor 
formulation for a homogeneous sphere (equation \ref{e:dopp-eq}) with 
$\theta_d \simeq 1.8\sqrt{\theta_{G_a}\theta_{G_b}}$. 
So for K1, $\theta_d = 0.55 \pm 0.02$ mas and we obtain
an equipartition Doppler factor for the {\em pattern} of
\begin{equation}
\label{e:equi-dopp}
\delta_{p} = 19.1^{+5.9}_{-2.9} \left(\frac{\eta}{D_A}\right)^{1/7}
(\delta'_f)^{-6/7}
\end{equation}
\noindent where $\eta = U_B/U_{rp}$ is the equipartition factor
($U_B =$ magnetic field energy density; 
$U_{rp} =$ energy density in the radiating particles), $D_A$ is the
angular size distance in Gpc, and $\delta'_f$ is the Doppler 
factor of the flow as viewed by an observer co-moving with the pattern.
To do this computation, we have assumed energy spectrum limits of 
$\gamma_1 = 10$ and $\gamma_2 = 1\times10^6$.  For $\alpha \simeq 0.5$,
the dependence on these limits is approximately $[\ln(\gamma_2/\gamma_1)]^{1/7}$.

\begin{figure}
\epsfig{file=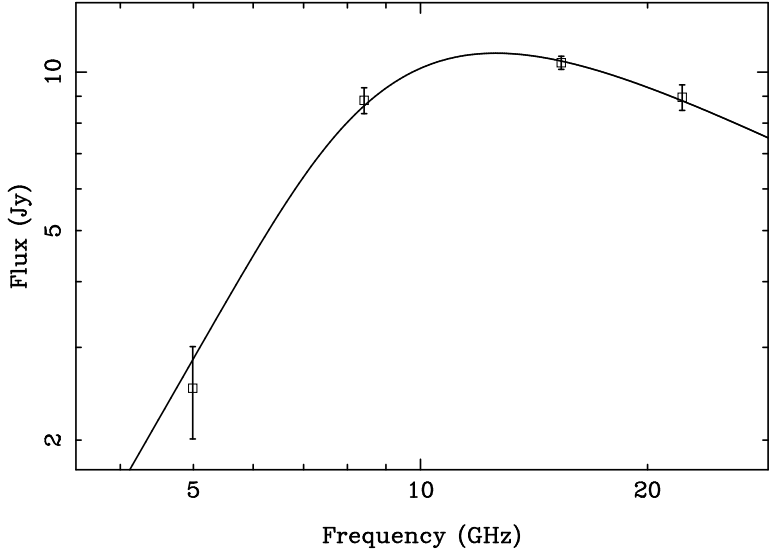, angle=-90, width=4in}
\caption{\label{f:spec.k4}
Synchrotron self-absorption spectrum of component K4. $\nu_{peak} = 12.59$ GHz 
$(+0.70,-0.41)$, $S_{peak} = 10.86$ Jy $(+0.40,-0.37)$, and 
$\alpha = 0.70$ $(+0.17,-0.16)$. 
}
\end{figure}

It is interesting to compare this Doppler factor for K1 to a Doppler
factor measured for the component K4.  The spectral fit for K4 is 
given in Figure \ref{f:spec.k4}, we find $\nu_{peak} = 12.59$ GHz 
$(+0.70,-0.41)$, $S_{peak} = 10.86$ Jy $(+0.40,-0.37)$, and 
$\alpha = 0.70$ $(+0.17,-0.16)$ with $\chi^2 = 0.8$.    
We have only an upper limit on the angular size of the component 
transverse to the jet direction, so we can only use the limit 
$\theta_d \leq 0.42 \pm 0.01$.  We calculate an equipartition 
pattern Doppler factor, 
\begin{equation}
\delta_{p} \geq 16.7^{+3.6}_{-2.1} \left(\frac{\eta}{D_A}\right)^{1/7.4}
(\delta'_f)^{-6.4/7.4}
\end{equation}
\noindent for the component K4. The dependence on the energy spectrum
limits for $\alpha \simeq 0.7$ is approximately $\gamma_1^{-0.4/7.4}$.
\citet{GPCM93} report an SSC Doppler factor for the core of 3C\,279 
of $\delta_{SSC} \geq 18.0$.  
 
\placefigure{f:spec.k4}

\subsection{Aspect Ratio}

K1 is a narrow component oriented perpendicular to its position angle.  At
15 and 22 GHz the component is well resolved in both directions, at 8 GHz
it is less well resolved, and at 5 GHz the component is unresolved along
the direction of the jet.  Assuming that the component is not oblique,
the observed aspect ratio is $\zeta = 0.43 \pm 0.03$.

Component K1 shows no sign of significant obliqueness.  At 22 GHz we
measure its orientation to be $7^\circ \pm 11^\circ$ from perpendicular
to its long term structural position angle.  The high frequency linear 
polarization of K1 is aligned with its long term structural position angle
to $-3^\circ \pm 6^\circ$. (The overall calibration of the polarization
position angle for our VLBA observations in Dec. 1997 was from simultaneous
VLA observations of the compact source OJ287.)  Using these estimates on the 
obliqueness in the plane of the sky as a guide, we estimate an uncertainty 
in the obliqueness in the plane of observation of $\pm 5^\circ$.  This 
uncertainty in the degree of obliqueness translates to an additional 
uncertainty in the measured aspect ratio, roughly $\zeta = 0.43 \pm 0.08$.

\subsection{Measuring the Distance}

Using our measurement of the observed proper motion, pattern Doppler factor, 
and aspect ratio of component K1, equation \ref{e:lower} gives the following
limit on the angular size distance to 3C\,279:
\begin{equation}
\label{e:dist-lower}
D_A \geq 1.8^{+0.5}_{-0.3} \eta^{1/8} ~~Gpc 
%D_A \geq 1.76^{+0.53}_{-0.30} \eta^{1/8} ~~Gpc
\end{equation}
\noindent which depends only on the equipartition factor, $\eta$.  Note that
we have used equation \ref{e:limit-df} and our measurement of 
$\zeta$ ($\geq |\cos\theta'|$) to limit the Doppler factor of the flow
relative to the pattern frame to $\delta'_f \leq 1.1 \pm 0.1$.  
(Because we have only an upper limit 
on $\delta'_f$, the upper limit on the distance (equation \ref{e:upper}) 
is undetermined.)  

\section{Discussion}
\label{s:sys-err}

Equation \ref{e:dist-lower} provides only a lower limit on the 
angular size distance to 3C\,279. The result is a limit because we 
could only make use of the aspect ratio constraint.  In general this technique 
can provide direct measurements (not just limits) for sources where 
some of the other constraints in Section \ref{s:constrain} can be 
applied successfully.  

The 1 $\sigma$ errors on this limit are $+28\%$ and $-17\%$.  These
errors are dominated by the uncertainty in the spectral turnover measurement
of component K1.  For K1, we have only one spectral point on the optically
thick side of the turnover and this point is poorly constrained due reduced
resolution at 5 GHz.  The spectral turnover for component K4 is better
determined giving a Doppler factor with 1 $\sigma$ errors of $+22\%$ and $-13\%$.
With better frequency coverage (perhaps by using widely separated IF channels near
the spectral turnover) and better angular resolution at lower frequencies 
(through the use of space VLBI), we believe we can eventually reduce the 1 $\sigma$
measurement errors on Doppler factors from equipartition and SSC 
techniques to $\sim  10-15\%$.  

For our 3C\,279 distance limit, the equipartition factor, $\eta$, is the most 
significant unknown quantity.  Even though $\eta$ enters to only a small factor, 
it is poorly constrained.  \citet{R94}, when proposing the
technique, argued that sources should be near equipartition 
($\eta = 1$ for electron-positron jets) and suggested
an error of $\sim 13\%$ in the Doppler factor for typical departures from
equipartition.  \citet{S86} calculated the diamagnetic
effect of spiraling electrons in a magnetic field.  He found that
the energy density of the electrons could not exceed 6 times the energy of the
magnetic field and still maintain synchrotron radiation.
\citet{BGT92} repeated Singal's calculation and
included a surface current term.  They found that the energy density of the
electrons could not exceed the energy density of the {\em applied} field by more
than a factor of 3 although they note that the energy density of the {\em effective} 
field in the region could be much smaller than the applied field.  

Another issue important to calculating precise values for equipartition
Doppler factors is the cutoffs in the power law particle energy spectrum.
The computed Doppler factors depend only weakly on the assumed value
of the cutoffs, but different, reasonable assumptions for the cutoffs
may lead to a $\sim 5-10\%$ uncertainty in the computed Doppler factor.
Energy spectrum cutoffs can be measured or constrained, however.  In 
\citet{WHOR98} we used VLBI circular polarization observations
to show the lower energy cutoff in 3C\,279 was $\gamma_1 \leq 20$.  The uncertainty
in the low energy cutoff dominates when $\alpha > 0.5$.

One way around the uncertainties in assuming equipartition and 
energy spectrum cutoffs is to calculate Doppler factors using measured 
x-ray fluxes.  A drawback to this approach is that the SSC Doppler
factor calculated for a given component is only a lower limit because, with
the capabilities of current instruments, observed x-ray fluxes include 
contributions from all parts of the parsec-scale source.  Such a limit is 
unlikely to be useful for jet components like K1 in 3C\,279 
which contribute a very small fraction of the total x-ray flux of the source.  
Assuming $\eta=1$ and inverting equation \ref{e:SSC-dopp} to solve for 
the x-ray flux of K1 yields $\simeq 0.01\mu$Jy of 2 KeV x-rays.  This
flux is less than $1\%$ of the total 2 KeV x-ray flux reported by \citet{W98}
for 3C\,279 in its quiescent state in January of 1996. 
It is interesting to note that the capabilities of the proposed MAXIM program 
(\url{http://maxim.gsfc.nasa.gov/}) would make x-ray observation of individual jet
components possible.  

The final area of systematic uncertainty is the assumed geometry for
the pattern.  We found that we obtained essentially identical spectral
fits when using the functional form for a uniform slab as for a uniform
sphere, so we can safely say that the assumed geometry has little 
affect on the spectral fit.  In appendix \ref{s:model-geom} 
we explore the remaining dependence of derived Doppler factors on assumed pattern geometry.
The main result is that a spherical geometry should be a good approximation
for calculational purposes and tends to produce a lower limit on the 
Doppler factor if the true geometry is non-spherical.

\section{Application to Cosmology}
\label{s:cosmo}

The angular size distance in terms of redshift, $z$, Hubble constant, $H_0$, 
matter density, $\Omega_M$, and cosmological constant, $\Omega_\Lambda = \Lambda/(3H_0^2)$ 
is given by, e.g. \citep{CPT92}
\begin{multline}
D_A = \frac{c}{H_0\sqrt{|\kappa|}(1+z)} \\
\times\mathbb{S}(
\sqrt{|\kappa|}\int_0^z [(1+z')^2(1+\Omega_Mz') \\
-z'(2+z')\Omega_\Lambda ]^{-1/2}dz')
\end{multline}
%\begin{equation}
%D_A = \frac{c}{H_0\sqrt{|\kappa|}(1+z)}\mathbb{S}\left(
%\sqrt{|\kappa|}\int_0^z \left[(1+z')^2(1+\Omega_Mz')-z'(2+z')
%\Omega_\Lambda\right]^{-1/2}dz'\right)
%\end{equation}
\noindent where if
$\Omega_M + \Omega_\Lambda > 1$ then $\mathbb{S}(x) = \sin(x)$ and  
$\kappa = 1-\Omega_M-\Omega_\Lambda$, if 
$\Omega_M + \Omega_\Lambda < 1$ then $\mathbb{S}(x) = \sinh(x)$ and 
$\kappa = 1-\Omega_M-\Omega_\Lambda$, and if
$\Omega_M + \Omega_\Lambda = 1$ then $\mathbb{S}(x) = x$ and 
$\kappa = 1$.

\begin{figure}
\epsfig{file=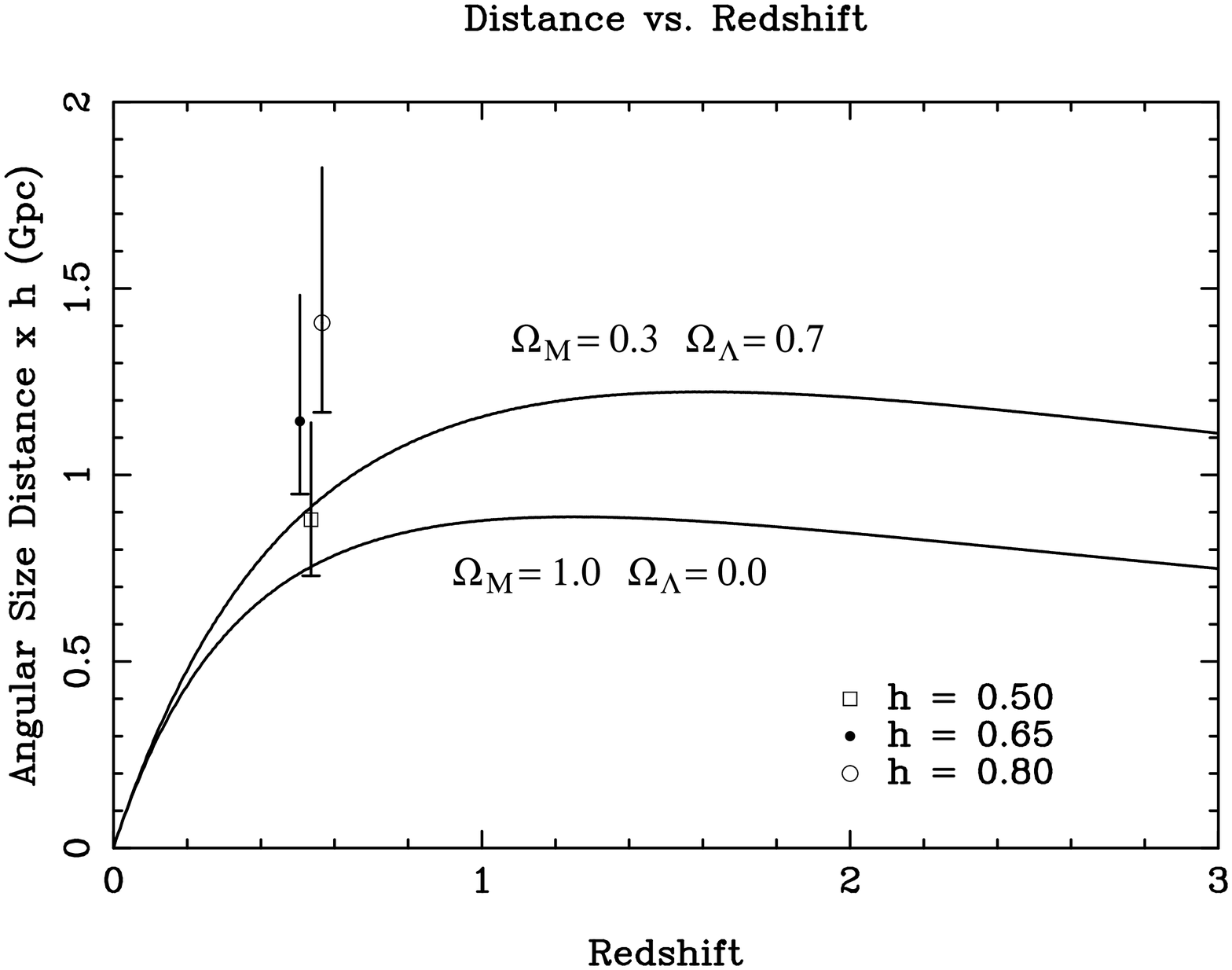, width=4in}
\caption{\label{f:dist}
Angular size distance versus redshift for two cosmological models for a flat 
universe ($\Omega_M+\Omega_\Lambda = 1$). The $\Omega_M = 0.3$, $\Omega_\Lambda=0.7$
case
is consistent with recent SNe Ia results, and the $\Omega_M = 1.0$, $\Omega_\Lambda=0.0$ 
case
corresponds to an Einstein-de Sitter universe.  The Hubble constant is taken to 
be $H_0 = 100h$ km/s/Mpc.  Our distance limit to 3C\,279 (assuming $\eta = 1$) is 
plotted for various values of $h$.  The points for $h = 0.65$ and $h = 0.80$ are 
offset slightly in redshift to enhance readability.
}
\end{figure}

In figure \ref{f:dist} we plot our lower limit on the distance to
3C\,279 (assuming $\eta = 1$) 
against two cosmological models for a flat universe ($\Omega_M+\Omega_\Lambda = 1$).  
The first case is consistent with recent type Ia supernovae results 
\citep{P99,R98}.  The second case is the
standard Einstein-de Sitter universe.  The Hubble constant is maintained 
as a free parameter which scales the ordinate of figure \ref{f:dist}.

\placefigure{f:dist}

Our single distance limit cannot distinguish between these 
cosmological models; however, we can investigate whether 
this technique holds promise for distinguishing these cases in the 
future.  At redshifts larger than $1.0$ these cosmological models
differ by as much as $30-40$\%.
Superluminal radio sources are regularly observed at these large redshifts.  
For 3C\,279 our lower limit has 1 $\sigma$ measurement errors on the order 
of $20-25\%$ and we believe it is reasonable obtain distance measures 
(or limits) good to $\sim 10-15\%$ for carefully planned observations on 
well selected objects.  Even a handful objects (over a range of redshift values) 
could provide strong constraints on cosmological models. 

The systematic uncertainties (discussed in \S{\ref{s:sys-err}}) in measuring the Doppler 
factor present the largest difficulty here.  We can constrain the range of
allowed values for the equipartition factor, $\eta$, by comparing distance 
measurements to sources at similar redshifts.  To detect any systematic 
offset in $\eta$ from unity, we
must have a way of calibrating our Doppler factor measurements.  Fortunately,
all of the poorly constrained quantities ($\eta$, energy spectrum cutoffs, geometric dependence)
can be grouped as a single multiplicative parameter in the equipartition Doppler factor.
We can calibrate any systematic offset in this parameter by making distance 
measurements to sources at moderate redshifts (where the effects of differing cosmological
models are not strong) and comparing the result to other distance measurement techniques.  
Such a scheme would introduce a dependence on the cosmological distance 
ladder, so we should actively seek other techniques for calibrating systematic 
effects in Doppler factor measurements.

\section{Conclusions}
\label{s:conclude}

We have demonstrated a new technique for directly measuring the distances
to high redshift, superluminal radio sources.  This technique involves the
comparison of Doppler factor and proper motion measurements for individual
jet components; we must also determine the jet angle to the
line of sight and pattern versus flow velocity.  We have presented several
techniques for measuring or usefully constraining these parameters.  In general
these techniques will be applicable only to selected sources which have jet
components with the right characteristics; however, with hundreds of
currently known superluminal sources and new jet components emerging
frequently from many of them, it seems reasonable to assume that we
will find a significant number of candidates.  One interesting
possibility is that some sources could have more than one component to
which we can apply these techniques, giving us multiple, independent
distance determinations to the same object.

To begin to usefully constrain cosmological parameters we 
need to obtain high quality distance measurements or limits to several 
sources over a range of redshift.  We have performed a detailed analysis of 
the measurement error associated with our limit on the distance to 3C\,279.  
We found the measurement error was $\sim 20-25$\% and concluded that carefully 
planned, high-quality observations could reduce measurement error to $\sim 10-15$\%.  
The systematic error was more difficult to quantify, although we outlined the 
sources of systematic error and presented rough estimates.  Applying this technique 
to a larger sample of sources will be important not only for probing
cosmological parameters but also for investigating the sources of systematic error
and how close these sources are to equipartition.

\section{Acknowledgments}

This work has been supported by NASA Grants NGT-51658 and NGT5-50136
and NSF Grants AST 92-24848, AST 95-29228, and AST 98-02708.

%%%%%%%%%%%%%%%%%%%%%%
%%                  %%  
%%    Appendix      %%
%%                  %%
%%%%%%%%%%%%%%%%%%%%%%

\appendix

\section{Doppler Factor Formulae}

While the relativistic plasma in radio jets does not contain atoms and molecules
whose characteristic spectra we could use to directly measure their bulk Doppler
factors, they do contain a unique (but broad) spectral feature: the synchrotron
self-absorption turnover.  The location of the turnover due to synchrotron 
self-absorption depends on the magnetic field strength, particle density, 
and size of the emission region.  For a volume of homogeneous plasma, we can
use the extrapolated optically thin flux at the turnover frequency and the 
optical depth at the turnover frequency to solve for the magnetic field and the
particle density, e.g. \citep{M87}. 

\begin{equation}
B = \frac{\delta_p\delta'_f}{(1+z)}C_{3} \tau_m^2 \nu_m^5 S_m^{-2} \Omega^2 
\end{equation}
\noindent and
\begin{equation}
K = \frac{(1+z)^{4+2\alpha}}{\delta_p^{4+2\alpha}(\delta'_f)^{3+2\alpha}}
    C_{4} \tau_m^{-(2\alpha+2)} \nu_m^{-(4\alpha+5)} S_m^{2\alpha+3} 
    \Omega^{-(2\alpha+3)}(D_A\xi_c)^{-1}
\end{equation}
\noindent where 
$K$ is the constant in the power law particle density,
$N_\gamma d\gamma = K\gamma^{-(2\alpha+1)}d\gamma$.  The sign of the spectral
index, $\alpha$, for optically thin emission is given by $S \propto \nu^{-\alpha}$. 
The parameter $\Omega = \left[\frac{V'}{D_A^2s_c'}\right]$ and $\xi_c = s_c'/D_A$, where   
$V'$ is the volume in the pattern frame and $s_c'$ is the line of sight
through the center of the volume along which the optical depth at the turnover, $\tau_m$,
is defined. The {\em extrapolated} optically thin flux\footnote{The factor by 
which $S_m$ over-predicts the observed peak flux, $S_o$, is
tabulated in table \ref{t:constants} for a spherical geometry.} 
at the turnover frequency, $\nu_m$, is given by $S_m$.  
Constants $C_3$ and $C_4$ are tabulated in 
table \ref{t:constants} for $\Omega$ in mas$^2$, $\nu_m$ in GHz, $S_m$ in Jy, $D_A$ in Gpc, 
and $\xi_c$ in mas.

Assuming we can translate the source dimensions, $V'$ and $s_c'$, into observable
quantities (e.g. observed angular size), these expressions provide us
with two equations and three unknowns, $B$, $K$, and the Doppler factor.
To solve for these quantities, it is necessary to have a third constraint.
Two possibilities for a third constraint are equipartition \citep{R94} and Synchrotron
Self-Compton (SSC) x-ray flux \citep{M87}.  

\subsection{Equipartition}

Equipartition assumes that the energy density of the magnetic field and the energy 
density of the particles are equal.  We will parameterize the relationship between 
the energy density of the magnetic field and the energy density in the 
{\em radiating} particles:
\begin{equation}
U_B = \eta U_{rp}
\end{equation}
\noindent where $U_B = B^2/8\pi$, 
$U_{rp} = \int mc^2\gamma N_\gamma d\gamma$.  For an electron-positron jet, 
$\eta = 1$ for equipartition.  In an electron-proton jet, the value of 
$\eta$ for equipartition will depend on the details of the particle 
acceleration within the jet. (See \S{\ref{s:sys-err}} for discussion
of the value of $\eta$.)

The equipartition condition is
\begin{equation}
B^2/8\pi = \eta mc^2 g(\alpha,\gamma_1,\gamma_2)K,
\end{equation}
\noindent where for $\alpha \neq 0.5$
\[ g(\alpha, \gamma_1 , \gamma_2 ) 
= \frac{1}{2\alpha - 1} \left(\gamma_1^{-(2\alpha-1)} 
- \gamma_2^{-(2\alpha-1)} \right) \]
\noindent and for $\alpha = 0.5$
\[ g(\alpha , \gamma_1 , \gamma_2 ) = \ln 
\left( \frac{\gamma_2}{\gamma_1} \right). \]  
\noindent $\gamma_1$ and $\gamma_2$ represent 
the lower and upper cutoffs for the particle energy distribution.  

Substituting the expressions for $B$ and $K$ into the equipartition 
condition and solving for the Doppler factor yields
\begin{equation}
\delta_{eq} = \delta_{p}(\delta'_{f})^{\frac{2\alpha+5}{2\alpha+6}} 
 = F_{eq} \tau_m^{-1} (1+z) 
   \left[ \frac{g(\alpha,\gamma_1,\gamma_2) \eta S_m^{2\alpha+7}} 
               {D_A\xi_c \Omega^{2\alpha+7}\nu_m^{4\alpha+15}}
   \right]^{\frac{1}{2\alpha+6}}
\end{equation}
\noindent where $F_{eq}$ is tabulated in table \ref{t:constants} for 
$\Omega$ in mas$^2$, $\nu_m$ in GHz, $S_m$ in Jy, $D_A$ in Gpc, 
and $\xi_c$ in mas.

\paragraph{Homogeneous Sphere Geometry:}
The spherical geometry assumes that the components emitted from the 
AGN are homogeneous spheres of radiating plasma. The sphere has a radius, $R$,
and we can define an angular diameter, $\theta_d = 2R/D_A$. 
Therefore $\Omega = (\pi/6)\theta_d^2$ and $\xi_c = \theta_d$.  With 
these identifications, the equipartition Doppler factor is given by
\begin{equation}
\label{e:dopp-eq}
\delta_{eq} = F_{eq} \tau_m^{-1} (1+z) 
\left(\frac{6}{\pi}\right)^\frac{2\alpha+7}{2\alpha+6}
\left[ \frac{g(\alpha,\gamma_1,\gamma_2) \eta S_m^{2\alpha+7}} {D_A (\theta_d \nu_m)^{4\alpha+15}}
\right]^{\frac{1}{2\alpha+6}}.
\end{equation}

The equipartition Doppler factor derived by \citet{R94} has a slightly different
functional form than our expression.  Readhead's expression is derived by comparing
observed brightness temperature, $T'_b$, to an (rest-frame) equipartition brightness 
temperature, $T_{eq}$: $\delta_{eq} = T'_b/T_{eq}$.  For 3C\,279 we can turn this expression
around and calculate $T_{eq}$ for component K1 from our measured $\delta_{eq}$
and observed brightness temperature.  We find $T_{eq} = 4\times 10^{10} K$ 
(for $\eta = 1$) which is close to $T_{eq} \sim 5\times 10^{10} K$ which
Readhead argues is a typical upper cutoff for powerful extra-galactic radio sources in 
their rest-frame.   
   
\subsection{Synchrotron Self-Compton Emission}
A second possible constraint is synchrotron self-Compton x-ray flux, 
e.g. \citep{M87}. The observed x-ray flux from the SSC process 
is given by (adapted from \citet{RL79})
\begin{equation}
S_X(\nu_c) = \Xi\frac{3\sigma_T}{8} KA(p)l'' S_m 
\left(\frac{\nu_m}{\nu_c} \right)^\alpha
                 \ln\left[\frac{\nu_b}{\nu_m}\right]   
\end{equation}
\noindent where 
\[ A(p) = 2^{p+3} \frac{p^2+4p+11}{(p+3)^2(p+5)(p+1)} \qquad p = 2\alpha+1 \]
\noindent and $\Xi$ is a factor ($\sim 1$) that accounts for the 
differences in photon number density (resulting from edge effects) within the 
emitting volume, 
$l''$ is the average line of sight as seen from the center of 
the geometry in the fluid frame, $\nu_b$ is the upper cutoff frequency of
the synchrotron emission spectrum, and $\nu_c$ is the x-ray 
observation frequency.  As defined earlier, $S_m$ is the extrapolated, optically
thin synchrotron flux at the turnover frequency, $\nu_m$.

Substituting for $K$ and solving for the Doppler factor gives
\begin{equation}
\delta_{SSC} = \delta_{p}(\delta'_{f})^\frac{2\alpha+3}{2\alpha+4} 
    = F_{SSC} S_m (1+z) \left[ \frac{\Xi l''}{s_c'}  
    \frac{ \ln\left[\frac{\nu_b}{\nu_m}\right] 
    \nu_m^{-(3\alpha+5)}\tau_m^{-(2\alpha+2)}} 
    { S_X (h\nu_c)_{KeV}^\alpha \Omega^{2\alpha+3} }
    \right]^\frac{1}{2\alpha+4}
\end{equation}
\noindent where $F_{SSC}$ is a constant which is tabulated in table \ref{t:constants} 
for $\Omega$ in mas$^2$, $\nu_m$ in GHz, $S_m$ in Jy, and $S_X$ in $\mu$Jy.

\placetable{t:constants}

It is important to note that, in practical application, the SSC Doppler factor 
is only a lower limit.  With the capabilities of current instruments, observed 
x-ray fluxes include contributions from all parts of the parsec scale jet, 
thermal x-ray emission from the accretion
disk, and inverse-Compton x-rays which are not the result of the SSC process.  
We therefore obtain only an upper limit on the SSC x-ray flux of a given 
component and thus a lower limit on its Doppler factor. 

\paragraph{Homogeneous Sphere Geometry:}
$\Omega = (\pi/6)\theta_d^2$ and $s_c' = 2R$.
Also, for a spherical geometry\footnote{This is strictly
true only if the pattern and flow are moving with the same speed.} $l'' = R$
and $\Xi = 3/4$ \citep{G79}.  
Making these substitutions gives the following expression
for the SSC Doppler factor:  
\begin{equation}
\label{e:SSC-dopp}
\delta_{SSC} 
    = F_{SSC} S_m (1+z) 
   \left(\frac{6}{\pi}\right)^\frac{2\alpha+3}{2\alpha+4} 
   \left[\frac{3}{8}
    \frac{\ln\left[\frac{\nu_b}{\nu_m}\right] 
    \nu_m^{-(3\alpha+5)} \tau_m^{-(2\alpha+2)}} 
    { S_X (h\nu_c)_{KeV}^\alpha 
    \theta_d^{4\alpha+6} }
    \right]^\frac{1}{2\alpha+4}.
\end{equation}

\section{Choice of Model Geometry for Doppler Factor Measurements}
\label{s:model-geom}

Differences between model geometries show up in 
the range of optical depths across the component (which affects both 
$\tau_m$ and the ratio $S_m/S_o$) and conversion of 
measured Gaussian FWHM diameters, $\theta_G$, to the angular dimensions of 
the assumed geometry.  These conversion factors can be estimated for
simple geometries by matching the second moment of their Fourier
transforms which are fit in the (u,v)-plane by Gaussian models.  For
a uniformly bright disk, $\theta_d \simeq 1.7\theta_G$.
For a homogeneous sphere, $\theta_d \simeq 1.9\theta_G$.

For the purposes of measuring the Doppler factor,
the effect of having a wide range of optical depths for a homogeneous sphere 
nearly offsets the larger conversion factor for Gaussian measured 
component dimensions.  If we assume that a given component is a sphere,
when in reality it is a uniformly bright disk, we will calculate a Doppler
factor that is about $10\%$ too small using the equipartition formula or
about $5\%$ too small using the SSC formula.  In this scenario
there is an additional factor by which assuming a sphere will under-predict
the Doppler factor.  This factor is due to the unknown physical depth of
the uniformly bright disk.  For the equipartition formula, this factor
comes in as $\sim (\theta_d/\xi)^{1/7}$, where $\xi$ is the angular
thickness of the disk.  For the SSC formula,  this factor
is $\sim\left[(8/3)(l''/s')\Xi\right]^{1/5}$, where $l''$ is the mean
line of sight as seen by a photon at the center of the disk (in the 
fluid frame), $s'$ is the physical depth of the disk (pattern frame),
and $\Xi \sim 1$ accounts for differences in 
photon density throughout the disk.

In general, the patterns we observe are likely to be some compromise 
between these geometries, perhaps a cylindrical disk viewed nearly
edge on.  Without detailed knowledge of the true geometry a 
spherical geometry is well-suited for calculation because it computes 
the angular area presented to the observer reasonably, 
gives a sensible line of sight through the component, and provides 
naturally for a range of optical depths across the component.  
\citet{M87} suggests using $\theta_d \simeq 1.8\sqrt{\theta_{G_a}\theta_{G_b}}$ 
to convert Gaussian FWHM dimensions to spherical diameters, 
and we have adopted his approximation.

%%%%%%%%%%%%%%%%%%%%%%
%%                  %%  
%%    References    %%
%%                  %%
%%%%%%%%%%%%%%%%%%%%%%

%%%%%%%%%%%%%%%%%%%%%%
%%                  %%  
%%     Figures      %%
%%                  %%
%%%%%%%%%%%%%%%%%%%%%%

\newpage
\onecolumn

%%%%%%%%%%%%%%%%%%%%%%
%%                  %%  
%%     Tables       %%
%%                  %%
%%%%%%%%%%%%%%%%%%%%%%

\newpage

  %%%%%%%%%%%%%%%%%
  %%             %%   
  %%   Table 1   %% 
  %%             %%
  %%%%%%%%%%%%%%%%%

\begin{table}
\begin{scriptsize}
\begin{center}
\tablenum{1}
\caption[]{\label{t:3c279t}Multi-frequency Component Data for 3C\,279 for epoch 1997.94.\\}
\begin{tabular}{cccccccccc}
\tableline \tableline 
$Component$ & $Frequency$ & $I$ & $m_L$ & $\chi$ 
& $R$ & $\Theta$ & $maj. axis$ & 
$min. axis$ & $\phi$ \\ 
&(GHz)& (Jy) & (\%) & (deg) & (mas) & (deg) & (mas) & (mas) & (deg) \\
\tableline 
D & $4.99$ & $5.20$ & $8.9$ & $-82$ & $\ldots$ & $\ldots$ & $\ldots$ & $\ldots$ & $\ldots$ \\ 
 & $8.42$ & $7.31$ & $7.4$ & $-46$ & $\ldots$ & $\ldots$ & $\ldots$ & $\ldots$ & $\ldots$ \\ 
 & $15.37$ & $12.08$ & $10.6\tablenotemark{a}$ & $-4\tablenotemark{a}$ & $\ldots$ & $\ldots$ & $\ldots$ & $\ldots$ & $\ldots$ \\ 
 & $22.23$ & $16.27$ & $8.2\tablenotemark{a}$ & $-10\tablenotemark{a}$ & $\ldots$ & $\ldots$ & $0.10$ & $< 0.15$ & $72$ \\ 
&&&&&&&&&\\ 
K4 & $4.99$ & $2.51$ & $3.1$ & $-28$ & $0.48$ & $-121$ & $\ldots$ & $\ldots$ & $\ldots$ \\ 
 & $8.42$ & $8.84$ & $2.9$ & $41$ & $0.33$ & $-126$ & $0.38$ & $< 0.50$ & $61$ \\ 
 & $15.37$ & $10.42$ & $10.9\tablenotemark{a}$ & $84\tablenotemark{a}$ & $0.32$ & $-132$ & $0.37$ & $< 0.25$ & $71$ \\ 
 & $22.23$ & $8.96$ & $10.2\tablenotemark{a}$ & $70\tablenotemark{a}$ & $0.35$ & $-135$ & $0.36$ & $< 0.15$ & $70$ \\ 
&&&&&&&&&\\ 
K2 & $4.99$ & $0.62$ & $17.5$ & $81$ & $2.88$ & $-123$ & $1.06$ & $< 0.30$ & $-17$ \\ 
 & $8.42$ & $0.29$ & $21.2$ & $88$ & $2.87$ & $-130$ & $2.08$ & $0.21$ & $-30$ \\ 
 & $15.37$ & $0.23$ & $17.4$ & $81$ & $3.12$ & $-124$ & $1.23$ & $0.27$ & $7$ \\ 
 & $22.23$ & $0.19$ & $11.8$ & $66$ & $3.07$ & $-125$ & $1.31$ & $< 0.07$ & $2$ \\ 
&&&&&&&&&\\ 
K1 & $4.99$ & $4.26$ & $4.2$ & $93$ & $3.21$ & $-114$ & $0.41\tablenotemark{b}$ & $< 0.30$ & $-24$ \\ 
 & $8.42$ & $4.17$ & $6.0$ & $76$ & $3.31$ & $-114$ & $0.46\tablenotemark{b}$ & $0.18\tablenotemark{b}$ & $-27$ \\ 
 & $15.37$ & $3.11$ & $5.9$ & $64$ & $3.36$ & $-115$ & $0.44$ & $0.19$ & $-19$ \\ 
 & $22.23$ & $2.68$ & $6.5$ & $63$ & $3.37$ & $-115$ & $0.46$ & $0.20$ & $-17$ \\ 
\tableline 
\end{tabular} 
\tablecomments{Limits on angular size are estimated to be $1/5$ of the uniformly 
weighted beam width along the corresponding dimension.}
\tablenotetext{a}{These values were obtained by allowing the position of the core, D,
to float in linear polarization and should be used with some caution.}
\tablenotetext{b}{These angular dimensions appear resolved in the model-fit but are somewhat
smaller than the formal limits.}
\end{center}
\end{scriptsize} 
\end{table} 

  %%%%%%%%%%%%%%%%%
  %%             %%   
  %%   Table 2   %% 
  %%             %%
  %%%%%%%%%%%%%%%%%

\begin{table}
\begin{scriptsize}
\begin{center}
\tablenum{2}
\caption[]{\label{t:constants} Tabulated Constants.\\}
\begin{tabular}{ccccccc}
\tableline \tableline 
$\alpha$ & $C_3$ & 
$C_4$ & $F_{eq}$ & $F_{SSC}$ & $\tau_m$ & $S_m/S_o$ \\ 
\tableline 
$0.1$ & $2.14\times 10^{-3}$ & $1.50$ & $1.36$ & $1.02$ & $0.104$ & $1.04$ \\
$0.2$ & $1.39\times 10^{-3}$ & $10.2$ & $2.08$ & $1.01$ & $0.204$ & $1.08$  \\
$0.3$ & $9.62\times 10^{-4}$ & $66.3$ & $3.02$ & $1.00$ & $0.300$ & $1.12$ \\
$0.4$ & $7.01\times 10^{-4}$ & $417$ & $4.20$ & $0.987$ & $0.392$ & $1.15$  \\
$0.5$ & $5.30\times 10^{-4}$ & $2.57\times 10^{+3}$ & $5.67$ & $0.972$ & $0.480$ & $1.19$ \\
$0.6$ & $4.12\times 10^{-4}$ & $1.56\times 10^{+4}$ & $7.44$ & $0.956$ & $0.565$ & $1.22$  \\
$0.7$ & $3.28\times 10^{-4}$ & $9.42\times 10^{+4}$ & $9.56$ & $0.942$  & $0.648$ & $1.26$ \\
$0.8$ & $2.66\times 10^{-4}$ & $5.65\times 10^{+5}$ & $12.1$ & $0.929$ & $0.727$ & $1.29$ \\
$0.9$ & $2.19\times 10^{-4}$ & $3.38\times 10^{+6}$ & $15.0$ & $0.918$ & $0.804$ & $1.33$  \\
$1.0$ & $1.82\times 10^{-4}$ & $2.02\times 10^{+7}$ & $18.3$ & $0.907$ & $0.878$ & $1.36$  \\
$1.1$ & $1.54\times 10^{-4}$ & $1.21\times 10^{+8}$ & $22.1$ & $0.899$ & $0.951$ & $1.39$ \\
$1.2$ & $1.31\times 10^{-4}$ & $7.26\times 10^{+8}$ & $26.4$ & $0.891$ & $1.02$ & $1.42$ \\
\tableline  
\end{tabular} 
\tablecomments{The optical depth at the turnover, $\tau_m$, is calculated 
for a homogeneous sphere.  The factor by which the extrapolated optically thin flux at
the turnover, $S_m$, over-predicts the observed peak flux, $S_o$, is also calculated for
the geometry of a homogeneous sphere.}
\end{center} 
\end{scriptsize} 
\end{table}

\end{document}